\documentstyle[12pt,aaspp4]{article}
\begin{document}
\title{A stratified framework for scalar-tensor theories of Modified Dynamics}
\author{R.H. Sanders}
\affil{Kapteyn Astronomical Institute, Groningen, The Netherlands}
\authoremail{sanders@astro.rug.nl}

\begin{abstract}

Although the modified dynamics (MOND) proposed by Milgrom successfully
accounts for the systematics of galaxy rotation curves and 
cluster dynamics without invoking dark matter, the idea remains a largely 
{\it ad hoc} modification of Newtonian dynamics with no basis in 
deeper theory.  Non-standard scalar-tensor theories 
have been suggested as a theoretical basis for MOND; however,
any such theory with the usual conformal relation between the Einstein
and physical metrics fails to predict the degree of light deflection
observed in distant clusters of galaxies.  The prediction is that there
should be no discrepancy between the detectable mass in stars and gas
and the lensing mass-- in sharp 
contradiction with the observations (Bekenstein and Sanders 1994).
In the present paper, I 
demonstrate that one can write down a framework for scalar-tensor theories 
which predict the MOND phenomenology for the low velocity ($v<<c$) dynamics
of galaxies and clusters of galaxies and 
which are consistent with observations of extragalactic gravitational  
lenses provided that one drops the requirement of the Lorentz Invariance
of gravitational dynamics.  This leads to  ``preferred frame'' theories
characterized by a non-conformal relation between the two metrics.  
I describe a toy theory in which
the local environment (the solar system, binary pulsars) is protected
from detectable preferred frame effects by the very same non-standard
(aquadratic) scalar Lagrangian which gives rise to the MOND phenomenology.
Although this particular theory is also
contrived, it represents a limiting case for two-field theories
of MOND, and is
consistent with a wide range of gravitational phenomena.  Moreover, it is a
cosmological effective theory which may explain the near
numerical coincidence
between the MOND acceleration parameter and the present value of the
Hubble parameter multiplied by the speed of light.

\end{abstract}

\section{Introduction}

More than ten years ago Milgrom proposed a modification 
of Newtonian dynamics (MOND) as an explanation of mass discrepancies in
astronomical systems with low internal acceleration (Milgrom 1983 a,b,c).  
As an alternative to dark matter, the idea has proven to be 
amazingly resilient in spite of sustained
attacks from several quarters (eg. The and White 1988, Lake 1989, 
Gerhard and Spergel 1992).  This is in large part due to 
the successes of the simple MOND prescription on a phenomenological
level, many of which were previewed by Milgrom in his original papers:
asymptotically flat rotation curves of spiral galaxies, the observed
form of the luminosity-velocity correlations for spiral and elliptical
galaxies (the Tully-Fisher and Faber-Jackson relations), the existence of
a critical maximum surface density in spirals and ellipticals (the Freeman
and Fish laws), the appearance of large mass discrepancies in low 
surface-brightness systems (eg. dwarf spheroidals and low surface-brightness
spirals),  the magnitude of
the discrepancy in clusters of galaxies, the magnitude of Virgo-centric
inflow.  

In addition, Milgrom used MOND to make rather specific
predictions which 
have been born out by subsequent observations.  There are two
notable examples: The first is 
the prediction that the rotation curves in luminous high
surface brightness galaxies should decline to an asymptotically flat value
while the rotation curve in low-luminosity, low surface brightness galaxies
should slowly rise to the asymptotically flat value (Milgrom 1983b).  
This effect has
been subsequently observed by Casertano and van
Gorkom (1991).  Secondly there was the suggestion, on the basis of high
MOND mass-to-light ratios of clusters of galaxies, that hot X-ray emitting
gas may make a very substantial contribution to the total observable mass
of clusters (Milgrom 1983c).  This has now been well-established by
ROSAT observations (B\"ohringer et al. 1993);  indeed, the predicted
MOND mass agrees remarkably well with the observed hot gas mass for a
number of clusters (Sanders 1994).

But perhaps the most outstanding success of MOND has been in connection
with the observed extended rotation curves of spiral galaxies.  It is
not that MOND, in some general sense, predicts flat rotation curves;  the
simple MOND formula predicts the precise form of the rotation
curve of a spiral galaxy from the observed distribution of stars and gas.
In a sample of galaxies with well-observed gas kinematics, there is
remarkable agreement between the observed and predicted rotation curves
using a single value of the 
MOND acceleration parameter $a_o$ (Begeman, et al. 1991, Sanders 1996).  
At the very least one can say that the MOND prescription provides
a far more economic fitting algorithm for galaxy rotation
curves than do multi-parameter dark halo models.      

In spite of these successes, the idea
is not yet taken seriously by most physicists and
astronomers.  The reason for this, to some extent, is the absence of
a solid theoretical underpinning of the idea;  MOND remains an
{\it ad hoc} empirically motivated modification of Newton's law
without connection to a more familiar theoretical framework.
There have been several attempts at writing down a relativistic (i.e.,
generally covariant) theory which reduces to MOND in the limit of low
accelerations (Bekenstein and Milgrom 1984, hereafter BM; Bekenstein
1988; Sanders, 1986, 1988; Romatka 1992), but these theories all contain
anomalies, or they are inconsistent with the classical local gravity
tests.  Moreover, they are 
designed to reproduce MOND in the low force limit, but are not based
upon some more general principle in the spirit of General Relativity (GR)
or modern gauge theories of particle physics.

There is one aspect of MOND which renders the idea less
{\it ad hoc} and which would seem to point the way to a more 
substantial theoretical basis.  That is the near 
numerical coincidence between
the empirically derived acceleration parameter, $a_o$, and $cH_o$,
the Hubble parameter multiplied by the speed of light.  This appears
to give a cosmological significance to the acceleration parameter,
and the implied relation between local dynamics and the expansion
of the Universe seems distinctly Machian (Milgrom 1983a, 1994).  
With respect to the theoretical
basis of MOND, this numerical coincidence suggests that the proper
theory should be an {\it effective} theory; that is to say, the MOND
phenomenology would only be predicted when the theory is considered
in a cosmological background.  Upon reflection it is evident that such
an effective theory cannot be provided by GR
because, in the context of GR, there is no cosmological
influence of this order on local gravitational physics.  However, various 
scalar-tensor theories do offer the possibility of such a connection.

Two types of scalar-tensor theories have been suggested to provide a
theoretical basis for MOND:  the so-called ``aquadratic Lagrangian''
(AQUAL) theories (BM, Sanders 1986) and a general class of two-scalar
theories of which ``phase coupling gravity'' (PCG) is the most discussed
example (Bekenstein 1988; Sanders 1988,
1989; Romatka 1992).  The AQUAL theories suffer, unavoidably, from
causality violations (super-luminal propagation of scalar waves) if the
MOND phenomenology is reproduced (BM, Bekenstein 1988), and PCG, apparently
in any form, provides no stable background solution for the two additional
fields (Bekenstein 1992).  But a more practical problem with these,
and indeed with all scalar-tensor theories in which the scalar field enters
as a conformal factor multiplying the Einstein metric, is the failure 
to predict the gravitational deflection of light at the level apparently
observed in rich clusters of galaxies (Bekenstein and Sanders 1994).
That is to say, if one wishes to replace dark matter by a modified
theory of gravity of the standard scalar-tensor type, then the scalar
field produces no enhanced deflection of light;  the observed deflection
would only be due to the detectable matter implying that the mass
of a system determined via 
gravitational lensing in the context of GR
should be substantially less than the conventional
virial mass.  This apparently is not the case (Fort and Mellier 1994).

With a view toward resolving the light-bending problem for scalar-tensor
theories, Bekenstein and Sanders (1994) considered a more general
relation between the Einstein and physical metrics-- the {\it disformal}
transformation, a relation which
includes both the scalar field and its gradient (Bekenstein 1992).  
The result was discouraging:  if the
propagation of classical gravitational waves is to be causal, then the
sign of the disformal term must be such that the light bending is 
actually reduced over that predicted by GR.

However, it now appears that the form of the transformation considered by
Bekenstein and Sanders is not the most
general relation between the gravitational and physical metrics, and that
a clue to the more general transformation is suggested by a class of 
theories known historically as ``stratified" theories (Ni 1972).  
Here, the physical metric is
related to the Einstein metric via a conformal factor involving
the scalar field and additional terms usually constructed from a 
non-dynamical vector field.  In some preferred frame, assumed to be
the cosmological frame, the vector has only a time component and 
space-like strata of the physical and Einstein metrics are conformally
related (historically, the ``Einstein'' metric is assumed to be 
the Minkowski metric so the theory is conformally flat on space-like
strata, but we will not make that restriction in the definition of
stratified theories considered here).  In stratified theories
the light bending in the weak-field limit can be 
equivalent to that predicted by 
GR; in fact, the original motivation for such theories
was to overcome the absence of light bending predicted by 
conformally flat scalar theories of gravity such as that of Nordstr\"om
(Misner, Thorne and Wheeler 1973). 

Aesthetically, such theories may be criticized because,
unlike GR, they contain {\it a priori} elements
such as a non-dynamical vector field and, in their original form, 
prior geometry described by the Minkowski metric.  This is certainly
contrary to the spirit if not the letter of General Covariance.  
Because they give a special status to the cosmic frame, such theories,
philosophically, go some way back toward the 
pre-relativity concepts of absolute space and time.
But aesthetics aside, the earlier stratified theories are all ruled
out because they predict 
various local preferred frame effects (such as earth tides) at a level
which should have been detected (Will 1993).

In the present paper I resurrect the idea of stratified theories
to provide a framework for scalar-tensor
theories of MOND in which the deflection of light bears the same relation
to the weak field force as in GR.  To achieve this I introduce
an additional vector field, here assumed to be non-dynamical,
into the matter Lagrangian in the form of a stratified theory.
But an additional element is that the vector field is also introduced
into the scalar field Lagrangian to form a new invariant which becomes the
square of the scalar cosmic time derivative ($\dot\phi^2$) in the preferred
cosmological frame (also an aspect of the generalized stratified theory
of Lee et al. 1974).  This allows one to write a cosmological effective
theory of MOND; i.e., one in which the acceleration parameter
$a_o$ is not put in by hand but is identified naturally with the
cosmic time derivative of the scalar field.

I describe a toy theory in which the total attraction is
inverse-square to high precision in the high acceleration limit
(e.g. near the sun) but the phenomenology is basically that of MOND
in the low acceleration limit.  In this particular example, 
the scalar field Lagrangian is
of the unconventional aquadratic or AQUAL form, although two-scalar
theories like PCG are also possible.  However, unlike AQUAL, PCG or any
scalar-tensor theory with a conformal relation between the two metrics,
this theory produce gravitational deflection of light at the level
predicted by GR with dark matter.
Locally (i.e., in the solar system) the stratified aquadratic
theories are weakly coupled
(but not arbitrarily weak) scalar-tensor theories, but unlike traditional
scalar-tensor theories (e.g. Brans-Dicke) the predicted deflection of light
about the sun is identical to that in GR.
And, unlike the traditional stratified theories,
the predicted local preferred frame effects may be suppressed by a large
(but not arbitrarily large) factor;  in fact, the 
current experimental limits on preferred frame effects are already at or near
the minimum level predicted by this theory.
There is, moreover, one additional predicted effect which might well
be observable:  a secular cosmic variation in the constant of gravity.

The basic goal here is to write down a 
relativistic theory of MOND, however contrived, in which the cosmological
background determines the value of $a_o$ and which is consistent with local 
and extragalactic phenomenology-- in particular with the observed
deflection of light by clusters of galaxies.  It would seem to be important 
to demonstrate that 
this is possible, particularly in view of the negative result of
Bekenstein and Sanders (1994) on conformally or disformally coupled
scalar fields.  The principal
conclusion is that  {\it stratified scalar-tensor} theories,
with a non-standard scalar field Lagrangian, 
can be consistent 
with the observations galaxy rotation curves and cosmic gravitational lenses 
as well as, at the present
levels of experimental precision, local gravitational dynamics.

\section{The effects of cosmology on local gravitational dynamics.}

In the context of GR, cosmology has an insignificant effect
on local gravitational dynamics.  Israelit and Rosen (1990) considered
the equation of motion of a particle in a cosmological background and
demonstrated that any additional acceleration on a particle orbiting
a galaxy at distance $r$ is on the order of $r{H_o}^2$; i.e., of the
same magnitude as the effect of a 
possible cosmological constant.  Therefore, MOND effects
which are postulated to be present on galactic scales at accelerations 
of $cH_o$, cannot possibly arise due to the influence of cosmology in
the context of pure GR.  

The general arguments for this
absence of significant cosmological effects were elucidated earlier 
by Will and Nordtvedt (1973) and are paraphrased below:  
Consider a local gravitational system
(the solar system, the Galaxy, a cluster of galaxies) embedded in
the Universe.  How does cosmology effect the gravitational physics of
this local system?  We divide the solution of a set of field equations
into two parts-- a cosmological solution and a local solution.  From
this viewpoint, cosmology establishes boundary conditions for the various
fields generated by the local system; i.e., the local system ``feels'' the
cosmology via its asymptotic field values.  Now the cosmological 
metric is the Robertson-Walker metric which, on scales small compared
to $c{H_o}^{-1}$ and short compared to ${H_o}^{-1}$ is the Minkowski metric.
In GR the metric field ($g_{\mu\nu}$) is the only field; 
therefore,
the local field must become Minkowskian, that of empty space, 
far from the mass concentration.
From the Birkhoff theorem we know that a spherically symmetric gravitational
field in empty space must be static with a metric given by the 
Schwarzschild solution.  This means that weak-field gravity is Newtonian
(G is unaffected by the presence of matter),
that there is no cosmic time dependence of G (local gravitational physics
is time-reversible) and that, due to the invariance properties of the
Minkowski metric, there are no preferred frame effects for systems in
uniform relative motion.

In standard scalar-tensor theory there are two fields ($g_{\mu\nu}$ and
$\phi$), and, as the theory is usually written, the scalar interacts with
matter jointly with $g_{\mu\nu}$ via a conformal transformation of the 
metric; i.e., the form of the interaction Langraian is taken to be
$$ L_I = L_I({\psi(\phi)}^2 g_{\mu\nu}...) \eqno(1)$$
where $\psi$ is a function of the scalar field.
In empty space, far from the mass concentration, the physical
metric is conformally flat.  The conformal function can be factored and
appears as a time-variable constant of gravity or universal mass function.
Thus the boundary condition on $g_{\mu\nu}$ 
remains Minkowskian and there are no preferred
frame effects.  Moreover, in standard scalar-tensor (i.e., Brans and
Dicke 1964,
Nordtvedt 1970, Wagoner 1968), with the usual quadratic Lagrangian
($\phi_{,\alpha}\phi^{,\alpha}$), the 
scalar ``force'' is essentially Newtonian and scales as the mass.  The
variation of $\phi$ with position is generally small compared to the
cosmological value (insignificant variation of G near a mass concentration),
but G is a function of cosmic time.  Therefore, gravitational physics is
not time reversible.

A theory such as Bekenstein's (1988) ``phase coupling gravity'' is a 
two-scalar-plus-tensor theory and, as such, offers the possibility of a more
dramatic cosmological effect on local dynamics.  Here, of the two scalar
fields $q$ and $\phi$, only one ($\phi$) couples to matter (jointly 
with $g_{\mu\nu}$ as in eq. 1) and the two fields interact via the 
kinetic term
of the matter-coupling field , i.e., the scalar field Lagrangian is
given by
$$L_s = {1\over 2}(q^2 {\phi_{,\alpha}}{\phi^{,\alpha}}+q_{,\alpha}
q^{,\alpha}). \eqno(2)$$  
Thus, in the field equation $q^2$ appears 
with respect to the gradient of $\phi$ in a form analogous to the
to the MOND function $\mu$ in the BM field equation (eq. 6a below); i.e.,
$$(q^2\phi^\alpha)_{;\alpha} = {{4\pi\eta G}\over{c^4}}T. \eqno(3)$$
where $\eta$ is a dimensionless parameter describing the strength of the
scalar field coupling and T is the contracted energy-momentum tensor
as usual.  Cosmology sets the
asymptotic value of $q$ which may be very different from its value 
near a local mass concentration.  This implies that local gravitational
physics may be strongly non-Newtonian (the scalar force is not necessarily
$1/r^2$ nor is the coupling to mass necessarily linear) which would seem
to be ideal for an effective theory of MOND.
Indeed it has been demonstrated (Sanders 1989) that with an appropriately
chosen, but somewhat un-natural, self-interaction potential for the scalar
field the predicted phenomenology is basically that of MOND.  The
scalar force exceeds the usual Newtonian force at accelerations below
an $a_o$ which is identified with the cosmic time derivative of $\phi$.
At accelerations much below $a_o$ the scalar force also becomes inverse
square but exceeds the Newtonian force by a factor $\delta$ (typically
assumed to be 10).  It can be shown that
$$a_o = 1.5\delta{\Omega_o} c H_o (t_oH_o)  \eqno(4) $$
where $\Omega_o$ is
the usual density parameter of the Universe and $t_o$ is the age
of the Universe.  It is
evident that with PCG in a cosmological setting, the MOND coincidence
is explained.

However, PCG has two serious failures:  the first concerns 
solar systems dynamics.  On the scale of the solar system PCG may be
considered as a Brans-Dicke theory with a weakly variable Brans-Dicke 
parameter $\omega$.  The problem is that in the form of the theory 
described by Sanders (1989)
$\omega$ in the solar system is much smaller than the experimental
lower limit of about 1000.  While it may be possible to avoid this problem 
choosing a different form for the self-interaction potential, the second
failure is more fundamental.  As noted in the Introduction, every
scalar-tensor theory in which the interaction with matter is described by
eq.\ 1, the scalar field has no
effect on the motion of photons and therefore could not explain the
enhanced deflection apparently observed in clusters of galaxies (Fort
and Mellier 1994) and, 
possibly, in individual galaxies (Brainerd, Blandford and Smail 1995).
Therefore PCG in its original form cannot be a viable theoretical 
replacement for dark matter.

A second scalar-tensor theoretical framework for MOND is provided
by theories with aquadratic Lagrangians (AQUAL) for the kinetic term
of the scalar field; i.e.,
$$L_s = {1\over 2}F(X){{{a_o}^2}\over{c^4}} \eqno(5a)$$
where $$X = {{\phi_{,\alpha} \phi^{,\alpha} c^4}\over {a_o}^2} \eqno(5b)$$ 
and F(X) is an arbitrary positive function of its unitless argument.
Combined with eq.\ 1 this leads to the BM field equation:
$$(\mu \phi^{,\alpha})_{;\alpha}={{4\pi G T}\over {c^4}} \eqno(6a)$$
with $$\mu = dF/dX = F'(X) \eqno(6b)$$
To yield MOND phenomenology $F'(X)$ must asymptotically approach $\sqrt{X}$ 
in the limit of small X.  

The original AQUAL theory of BM, as well as its trivial revision by
Sanders (1986), is in no sense a cosmological effective theory.  The
acceleration parameter is written in by hand, and the theory has no
obvious cosmological extension (as originally 
written, the Lagrangian becomes
imaginary if the scalar 4-gradient has only a time-component).  However,
by making use of an additional field, a non-dynamic vector field,
it is possible to write down an AQUAL theory in which the cosmic 
time derivative
of the scalar field, $\dot\phi$, can be inserted separately into the
scalar field Lagrangian.  
Since, in the appropriate units, $\dot\phi$ has a current value on
the order of $cH_o$, this provides an obvious mechanism for a cosmologically
imposed critical acceleration on the order of $a_o$.  
Moreover, given the vector field, such an AQUAL theory
can be written quite naturally as a stratified theory.  Then with 
the appropriate coupling of the 
scalar to the Einstein metric and to the vector field, the problem of
light-bending is solved.

\section{Stratified scalar-tensor theories and the deflection of
light.}

\subsection{A stratified theory including General Relativity}

In the historical stratified theories, several of which were 
designed to produce light bending equal to that
predicted by GR, the physical metric was
constructed from a non-dynamical vector field and the Minkowski metric, 
i.e., a prior geometry unrelated to the distribution of mass or 
energy (Ni 1972).  Here because we wish to retain 
GR in the strong field limit, we replace the Minkowski metric by the
Einstein metric.  Then, in the spirit of stratified theories, the 
physical metric, $\tilde g_{\mu\nu}$, is related to the
Einstein metric, $g_{\mu\nu}$, via the transformation
$$\tilde g_{\mu\nu}= u(\phi) g_{\mu\nu} - w(\phi) {A_\mu} {A_\nu} \eqno(7) $$ 
where $u(\phi)$ and $w(\phi)$ are at this point unspecified
functions of the scalar field $\phi$, and the dynamics of $g_{\mu\nu}$
is derived from the Hilbert action
$$S_g = {{c^4}\over {16\pi G}}{\int {R[g_{\mu\nu}]\sqrt{-g}}\, d^4 x}; 
\eqno(8)$$
i.e., the theory includes GR and, for u=1, w=0, reduces to GR.
We specify that $A^{\mu}$ is a non-dynamical
vector field with the only non-zero component pointing in the positive
time direction in the cosmic frame and tuned to the Einstein metric such that
$$g^{\mu\nu}A_\mu A_\mu = -1 \eqno(9)$$ (in some
theories $A_\mu = t_{,\mu}$ where $t$ is a non-dynamical cosmic time scalar).  
Thus, any frame at rest with respect to the Universe (i.e., the cosmic 
background radiation) becomes a preferred frame where the 
theory takes its simplest form.  

A word is required about the definition of a non-dynamical vector field
in a theory in which spacetime
is not {\it a priori} Minkowskian.  In a spacetime with
a high degree of symmetry (i.e., Robertson-Walker), the vector $A^\mu$ can
be uniquely defined as the unit vector orthogonal to space-like hyper-surfaces.
However, if we permit mass concentrations, as in the real Universe, 
the definition of a non-dynamical vector field becomes ambiguous as the
entire spacetime cannot necessarily be globally sliced by such surfaces 
(Bekenstein, private communication).  There are several possibilities for
removing this ambiguity, but it may be that a fully 
consistent theory requires
that the vector field be dynamical.  For now, because in problems of
galactic or solar system dynamics $g_{\mu\nu} \approx \eta_{\mu\nu}$
(with appropriately chosen physical units),
we assume that $A^\mu$ remains strictly time-like in any almost
Minkowskian frame at rest with respect to the cosmic frame.

With the coupling described by eq.\ 7, 
the particle action in the Einstein frame is given by
$${S_p} = -mc\int {\Bigl[-\{u(\phi) g_{\mu\nu} - w(\phi)A_{\mu}
A_{\nu}\}{{dx^\mu}\over{dp}}{{dx^\nu}\over{dp}}\Bigr]^{1\over 2}}dp  
\eqno(10)$$
where $p$ is some parameter along the path of the particle. 
Extremizing the action with respect
to variations in $x^\mu$ in the usual way and setting $dp = d\tau$ (the 
invariant interval), we find the covariant equation of motion:
$${{d{U_\nu}}\over{d\tau}} = {1\over 2} g_{\mu\kappa,\nu} U^{\kappa}
U^{\mu} u(\phi) + F_\nu  \eqno(11)$$
where $U_\nu = \tilde g_{\mu\nu}{{dx^\mu}\over{d\tau}}$ is the covariant velocity.
The first term on the right-hand-side is the usual Einstein-Newton
gravitational force and 
$F_\nu$ is the scalar force (in the Einstein frame the motion is
non-geodesic) given by
$$F_\nu =  {{\phi_{,\nu}}\over 2}{ \Bigl[-{u'\over u} + ({{u'w}\over u}
-w')  A_\kappa A_\mu U^\kappa U^\mu\Bigr]} \eqno(12)$$
where the prime indicates the derivative of $u$ or $w$ with respect to $\phi$. 

Because we are interested here in the equivalent Newtonian force (the
ordinary force or 3-force)
on slow or fast particles, we may also set $dp = dt$ (time in some specific
frame) in eq.\ 10 and rewrite the equation of motion as
$${{dp_i}\over dt} = -{1\over 2} m\,[g_{{oo},i} + g_{{jk},i} {v^j}{v^k}]
u(\phi) + F_i \eqno(13)$$
where $${p_i} ={mv_i}[-\tilde g_{oo} - \tilde g_{jk}{v^j}{v^k}]^{-{1\over 2}} 
\eqno(14)$$
is the  3-momentum and {\bf v} is the 3-velocity (the Greek indices
refer to spatial coordinates).  The first term on the right-hand
side again represents the Einstein-Newton force and $F_i$ is the ordinary
scalar force given now by
$$F_i = m{\phi_{,i}\over 2}{\Bigl[(u'g_{\mu\nu} - w'A_\mu A_\nu) {{dx^\mu}
\over
dt}{{dx^\nu}\over dt}\Bigr]}\Bigl[ -\tilde g_{\mu\nu}{{dx^\mu}\over dt}
{{dx^\nu}\over dt}\Bigr]^{-{1\over 2}}. \eqno(15)$$
In the weak field limit we may set $g_{\mu\nu} \approx \eta_{\mu\nu}$.
This, in effect, is defining the measure of time and length such that 
gravitational radiation propagates with unit velocity.  Then, in
the preferred frame where {\bf A} is strictly time-like,
we find for the ordinary scalar force
$$F_i = -m{{\phi_{,i}}\over 2}[u'+w'-u'v^2][(u+w)-uv^2]^{-{1\over 2}}
\eqno(16)$$
Dividing by the relativistic mass, 
$$m' = m u^{1\over 2}[(1 + w/u) -v^2]^{-{1\over 2}}\eqno(17)$$
we then determine the ordinary acceleration induced by the scalar field as 
$$f_i = -{\phi_{,i}\over 2}[u' + w' - u'v^2]u^{-1} \eqno(18)$$
where the particle speed $v$ approaches the limit 
$c' = \sqrt{1 + w/u}$ which is variable and may exceed one as measured in
units in which the Einstein metric is asymptotically Minkowskian. 
Setting $v=0$ gives the scalar acceleration on slow moving particles and
setting $v=c'$ gives the acceleration of relativistic particles or photons.
If we let $k$ be the ratio of the scalar acceleration on photons to that on
slow particles we find the simple result that
$$k = (w'-{{u'w}\over u})(u'+w')^{-1} \eqno(19)$$
That is to say, in the weak field
static limit, the deflection of a photon from a straight-line path would
be given by 
$$\theta = {2\over{c^2}}{\int{{f_n}^\bot} dz} + {k\over{c^2}}{\int{
f^{\bot}}dz} \eqno(20) $$
integrating along the path.  Here ${f_n}^\bot$ is the perpendicular 
component of the usual Newtonian acceleration (i.e., resulting from the
weak field limit of GR, the first term on the right-hand-side of eq.\ 13), 
and $f^\bot$ is the perpendicular component of
the scalar force on slow particles (eq.\ 16 with $v=0$).
For the usual scalar-tensor theory with a conformally coupled scalar field
$w(\phi) = 0$ which tells us immediately that $k = 0$; i.e., the scalar
field does not affect the motion of photons at all.  

\subsection{General constraints determine the free functions and light 
bending}

Several general physical considerations uniquely determine the form of 
the functions $u(\phi)$ and $w(\phi)$ and hence the relation of
the light bending to the weak field force (k in eq.\ 20).  First of all,
the condition that the physical and Einstein metric have the same
signature requires that $u(\phi)>0$.   
Further, it is reasonable to expect that physics should be
invariant to a global transformation of physical units, and
the appearance of a special direction in stratified
theories implies that units transformations may differ in
directions parallel and perpendicular to {\bf A} (Bekenstein, private
communication).  Representing
such a transformation as a shift in the zero of of $\phi$, as
in conformal theory, and considering coordinates in which $g_{\mu\nu}$
and $\tilde g_{\mu\nu}$ are diagonal, such invariance implies that
$$u(\phi) = e^{-\phi} \eqno(21)$$
and $$w(\phi) + u(\phi) = e^{\beta\phi} \eqno(22)$$

One more condition allows us to specify $\beta$ (following an argument
by Dicke 1957).
It is evident that the electro-magnetic invariant should 
contain the {\it physical} and not the gravitational metric, i.e.,
$$L_{em} = \tilde g_{\alpha\mu}\tilde g_{\beta\nu}F^{\alpha\beta}
F^{\mu\nu} \eqno(23)$$
because this yields trajectories for light corresponding to null
geodesics of the physical metric.  Then, in Maxwell's equations, the
effective dielectric parameter and permeability of empty space are given
by $$\epsilon = u(\phi) \eqno(24)$$
and $$\mu = {1\over {u(\phi) + w(\phi)}}.  \eqno(25)$$
in units such that the Einstein metric is asymptotically Minkowskian.
But the unitless physical constants, such as the fine structure 
constant $$ \alpha = {{e^2}\over h}{\Bigl({\mu\over\epsilon}
\Bigr)}^{1\over 2} \eqno(26)$$
should be independent of the choice of physical units.  Then it follows
that $\mu = \epsilon $ and $\beta=1$ in eq.\ 22. 
Thus we find $$w(\phi) =
e^\phi - e^{-\phi}. \eqno(27)$$  The functions u and w
have the form generally assumed in the traditional stratified theories
(Ni 1972).
Note that while light propagates along null geodesics of the physical metric,
gravitational radiation propagates along null geodesics of
the Einstein metric.  In units such that the physical metric is
asymptotically Minkowskian the velocity of propagation of gravitational
radiation becomes
$c_g = e^{-\phi}$.  Therefore causal propagation of gravitational 
waves requires that $\phi>0$; this should follow from any
sensible cosmology.

Given the form of $u(\phi)$ and $w(\phi)$ we find from eq.\ 19  
that $k=2$ as in the historical stratified theories;
i.e., the
weak field expression for the deflection of photons, eq.\ 20,  has the same 
relation to the total acceleration on slow particles as
in GR.  This has an immediate observational consequence:  any replacement
of dark matter by a stratified scalar-tensor theory yields light
bending exactly equivalent to that of GR plus dark matter; i.e., the 
lensing mass of a cluster determined by the usual formula
should be equal to that of the conventional viral mass. 

The overall conclusion of this section is that the use of the
non-conformal relation between the physical and Einstein metrics (eq.\ 7)
involving an additional cosmic field permits the scalar field to influence
the deflection of photons in a manner not anticipated by Bekenstein
and Sanders (1994).  Indeed, given very general constraints on the
frameworks free functions, the relation of the deflection angle to the
weak field force is the identical to that in GR.

\section{Aquadratic stratified theory}

\subsection{A generalized field action}

Having introduced the non-dynamical cosmological unit vector {\bf A} into
the general relation between the physical metric and the Einstein 
metric (eq.\ 7), we may also use it to form a second scalar field
invariant;  i.e., in addition to the usual invariant   
$$ I = g^{\mu\nu}\phi_{,\mu}\phi_{,\nu} \eqno(28a)$$
there is also $$J = A^{\mu} A^{\nu} \phi_{,\mu}\phi_{,\nu} \eqno(28b)$$
(here the theory will be written in the Einstein frame).
Because in the cosmological frame $A^\mu$ is postulated to be time-like, this
allows us to insert the cosmic time-derivative of $\phi$ directly into
the scalar field Lagrangian instead of introducing a new dimensional 
parameter, $a_o$ (as in the aquadratic theories of BM and 
Sanders 1986).  Moreover, if  $$K = I+J \eqno(29)$$
then K becomes square of the spatial gradient of $\phi$ in 
the preferred frame.  Thus we 
can manipulate the spatial and time derivatives independently in the preferred
frame already at the level of the field action.

The most general theory involving J and K is described by the action 
$$ S_{\phi} = {{c^4}\over{8\pi G}} \int {J\, Q(K/J)}\sqrt{-g}\,d^4x 
\eqno(30)$$ where Q(X) is any real function of its argument X.  
In particular, if $Q(X) = X-1$ we are left (in the preferred frame) 
with the usual quadratic scalar
field Lagrangian which, of course, yields an inverse-square attraction 
for the scalar force; the more general form yields 
AQUAL theories but with no new dimensional parameters.

The dynamics of the theory comes from the total action
$$S = S_g + S_\phi + S_m \eqno(31)$$
where $S_g$ is the gravitational action (eq.\ 8) and the matter action
$S_m$ is given by $S_p$ (eq.\ 10) summed over particles. 
Finding the extremum of the action with respect to $\phi$ 
gives the field equations
$$ {1\over \sqrt{-\tilde g}}(\sqrt{-g}P^{\alpha\beta} \phi_{,\beta})_{,\alpha} 
= {{4\pi G }\over c^4}  \tilde T^{\mu\nu}[g_{\mu\nu} e^{-\phi} + (e^{\phi} 
+ e^{-\phi})A_\mu A_\nu] \eqno(32a) $$
where
$$P^{\alpha\beta} = g^{\alpha\beta}Q'(X) + 
A^\alpha A^\beta [Q(X) + Q'(X) -XQ'(X)] \eqno(32b) $$
where $Q' = dQ/dX$.
In the preferred frame this becomes
$$P^{ij} = g^{ij}Q'(X)\,\,\,\, (i,j=1,2,3) \eqno(32c)$$
and
$$P^{tt} = Q(X)-XQ'(X) \eqno(32d) $$
The source is expressed in terms of the energy-momentum tensor
$$\tilde T^{\mu\nu}= -{2\over {\sqrt{-\tilde g}}}{\delta\over{\delta 
\tilde g^{\mu\nu}}} S_m
\eqno(32e) $$ in the physical frame so that the density $\rho$ and
pressure $p$ are those quantities actually measured by an observer.
This resembles the AQUAL field equation of BM except that the scalar function
of the invariant $\phi_{,\alpha}\phi^{,\alpha}$ ($F(X)$ in the notation of
BM, eq.\ 5a) has been replaced by a tensor $P^{\alpha\beta}$.
The complete theory includes the usual Einstein field equation for
$g_{\mu\nu}$ but with unconventional terms for the
contribution of the scalar field to the energy-momentum tensor.

Recalling that $X = {(\nabla \phi \cdot \nabla \phi)/\dot\phi^2}$ in the
preferred frame, 
let us choose $$Q(X) = F(X)- \kappa\eqno(33)$$ where $\kappa$ is a 
number on the order of one included to provide a cosmological solution,
and F is to be identified with the function of $(\nabla\phi/a_o)^2$
in the BM aquadratic theory (referred to below as the BM function).
There are no obvious {\it a priori} restrictions on the form of F, but
in order to reproduce both MOND phenomenology on the scale of galaxies
and precise inverse-square attraction in the solar system,
it must be the case that
$F(X) \propto X^{3\over 2}$ in the limit where $X<<1$ (see BM) and
$F(X) \propto X$ where $X>>1$.  This is because in the quasi-static 
case (no variation of $\phi$ on time-scales short compared to the Hubble time)
the MOND function is 
$$\mu(x) = F'(X) \eqno(34a) $$  (see eqs.\ 6a, 32a, 32c), as in BM theory
(eq.\ 6b),
has the appropriate asymptotic behavior (Milgrom 1983a). Here
$$x = \sqrt{X} = {|{\nabla \phi}/\dot\phi|}c \eqno(34b)$$
with the cosmic time derivative $\dot\phi$ playing the role of $a_o$
(here the speed of light is explicitly included to make the physical
units clearer below).  

\subsection{The cosmological origin of ${\rm a_o}$}

The identification of $\dot\phi$ with $a_o$ becomes evident when we consider, 
with eqs.\ 32 and 33,
the cosmic evolution of $\phi$.  With no spatial gradients 
$P^{tt} = -\kappa$ and eq.\ 32a becomes
$$e^{\phi}{d\over{dt}}(R^3 \dot\phi) = - {{4\pi G}\over \kappa}
(3p \,e^{-\phi} +\rho \,e^{\phi})R^3 \eqno(35) $$
where $R$ is the cosmic scale factor. 
It is clear from eq.\ 35 that pressure enters
as a source in the same way as density;  in particular, unlike standard
scalar-tensor theory, the radiation contributes to the source of $\phi$
at the level of twice the corresponding density for non-relativistic 
material.  This is nicely consistent with the result derived in the
previous section where we see that the scalar force produces twice as
much acceleration on photons as on slow-moving particles (see also 
Dicke 1957).  But in addition we see that in a pressureless universe
the evolution for $\phi$ is identical in form to that of a standard
conformally coupled scalar-tensor theory.  In particular we find
that $$\dot\phi = -{{4 \pi G\rho t}\over \kappa} \eqno(36)$$
where $t$ is cosmic time.  Here an integration constant has been arbitrarily
set to zero.  At the present epoch this
becomes $$\dot\phi_o = -1.5 {{\Omega_o}\over \kappa} {H_o}^2 t_o \eqno(37) $$
where $t_o$ is the present age of the universe.

Now consider the solution of the eqs.\ 32 and 33 about a point mass 
at rest in the preferred cosmological frame.  
This is simplified by assuming that the solution for $\phi$
about the mass concentration is quasi-static; i.e., there is no time
dependence on time-scales short compared to the Hubble time. 
Furthermore we assume that $\dot\phi_o$ has very weak
r-dependence and so appears as a constant in the spatial equation.
One then finds, from eqs.\ 32 and 33 that
$$x\mu(x) = {{GM}\over {r^2 c^2 |\dot\phi_o|}}\eqno(38) $$
Here a factor $e^{\phi}$ has been absorbed into
the definition of G.  We set $$\mu(x) = {1\over 2}k_cx \eqno(39)$$ 
in the limit where $x<<1$ (the required form in the MOND regime)
where $k_c$ is a number between 0.1 and
1 with a physical meaning to be described below.
Then in the low acceleration limit, eq.\ 38 becomes 
$$(\nabla \phi)^2 = {{2GM}\over{k_c r^2 c^3}} |\dot\phi|. \eqno(40)$$
From eq.\ 18, we find that the scalar force on slow particles is
$$f_s = {1\over 2} {\nabla \phi}\,c^2 \eqno(41)$$
or $$f_s = \Bigl({{GMc|\dot\phi|}\over{2k_cr^2}}\Bigr)^{1\over 2} \eqno(42)$$
This is identical to the MOND expression in the low acceleration limit
with $$a_o = {{c|\dot\phi}|\over{2k_c}} \eqno(43)$$
or, with eq.\ 37 $$a_o = {3\over{4k_c}}{\Omega_o\over{\kappa}}(t_o H_o) cH_o.
\eqno(44)$$ Thus the possibility of separating the time and space
gradients of $\phi$ in the preferred frame can provide
a cosmologically effective theory for MOND. 
  
\section{A limiting theory}

\subsection{Weak-field constraints on the Lagrangian}

Stratification can solve the light-bending problem of scalar-tensor
theories while providing a framework for cosmological effective
theories of MOND.  But can one
construct such a theory which is consistent both with the MOND
phenomenology and with local
gravitational dynamics?  In the weak field limit, scalar-tensor 
theories may be considered as two-field theories of gravity
where, in addition to the usual Newtonian force, $f_n$, there is a
scalar force, $f_s$ (a ``fifth'' force) which is given by eq.\ 38.
The simplest such aquadratic theory would be one in which the BM function is
$$F(X) = {1\over 3} X^{3\over 2} \eqno(45)$$ in eq.\ 33.
This yields a scalar force about a point mass with the form of eq.\ 42
(i.e., falling as 1/r) which exceeds the Newtonian force below 
accelerations of $a_o$; i.e. beyond a critical radius given by
$$r_c = \Bigl({{GM_\odot}\over{a_o}}\Bigr)^{1\over2}.\eqno(46)$$  
The total force in the solar system would
then be $$ f_\odot = {{GM_\odot}\over {r^2}} + {\Bigl({{GM_\odot a_o}
\over {r^2}}\Bigr)}^{1\over 2}. \eqno(47)$$ The problem is that the 
deviation of $f_\odot$ from inverse square attraction would severely 
violate the experimental constraints imposed by planetary precession
and limits on the variation of Kepler's constant, $K_\odot=GM_\odot$.
For example, in the outer solar system where the deviation would be the
largest, the predicted fractional variation in Kepler's constant at distance
r from the sun is $\Delta K_\odot/K_\odot = r/r_c$.  
At the orbit of Neptune this is
$4.2\times 10^{-3}$ (with $a_o=1.2\times 10^{-8}{\rm cm/s^2}$) 
which is more than 
a factor of 1000 times larger than the existing observational upper limit 
on $\Delta K_\odot/K_\odot$ ($\leq 2\times 10^{-6}$) between 
the orbit of Neptune and 
the inner planets (Anderson et al. 1995).

Therefore, a theory is required in which the total attraction
in the solar system is inverse square to very high precision while
yielding MOND phenomenology on the scale of galaxies.  A toy theory that
can meet these requirements is defined by 
$$F(X) = X/\eta \eqno(48a)$$
in the limit where $X\geq 1$ and 
$$ F(X) = {{1\over 3}{{k_c}} [1-{(X}^{3\over 2}]^{-1}} \eqno(48b)$$
where $X<1$ (the MOND limit). Here $\eta$ and $k_c$ are 
parameters of the theory (in the complete theory, eq.\ 33, it must
be that $\kappa > k_c$ for the existence of stable scalar waves).

With eqs.\ 34, 41, and 43 the MOND function becomes
$$\mu(x) = 1/\eta \,\,\,\,\,\,\, x>1 \eqno(49a)$$
$$\mu(x) = {1\over 2}{k_c x}{[1-x^3]}^{-2}\,\,\,\,\,\, x<1 \eqno(49b)$$
where the scalar force is given by $f_s = k_c x a_o$.  Therefore, $k_c$
is the transition acceleration due to the scalar field, in units of $a_o$, 
between the high and low acceleration limits of the theory. 

Eqs.\ 38, 41 and 49 may be solved numerically for the
the scalar force as a function of distance from a point mass, 
and this is shown in Fig.\ 1 in the case where $\eta = 1.25\times 10^{-5}$
and $k_c = 0.34$ (these values will be justified below).  
Here we see the Newtonian and scalar forces,
in units of $a_o$, as a function of radius, in units of $r_c$ (eq.\ 46).
The vertical dashed line corresponds to the position of the orbit of Neptune
assuming that the point mass is a solar mass and taking $a_o=1.2\times
10^{-8}{\rm cm/s^2}$ as implied by galaxy rotation curves (Begeman et al.
1991).  It is evident that the total
attraction ($f_n+f_s$) is inverse square to the orbit of Neptune;
at larger distance the scalar force remains nearly constant at the level
of $k_ca_o$ 
while the total acceleration decreases by about four orders of 
magnitude to a value near $a_o$; at still larger radii $f_s$ 
dominates the total attraction falling as 1/r.
It should be emphasized that there is no theoretical 
motivation for this assumed form of the the BM function (eqs.\ 48). 
It is the form required if the theory is to be consistent
with the inverse square 
law in the solar system while yielding modified dynamics at accelerations
below $cH_o$.

Eqs.\ 38, 41 and 49 may also 
be solved algebraically for $f_s$ about an extended spherically symmetric
mass distribution, where now $M$ is replaced by $M_r$, the enclosed
mass at radius r.  In Fig.\ 2, we see the predicted
rotation curves for several spherical galaxies with exponential density
distributions; the various values of the mass and length scale indicated.
It is evident that the curves are asymptotically flat and structureless
with the asymptotic velocity scaling as $M^{1\over 4}$ as in MOND.
However, if $k_c$ is smaller than about 0.3, rotation curves first decline
before rising to the asymptotic flat value, in contradiction to the 
observed form.

\subsection{Post-Newtonian constraints on the parameters of the theory}

Consistency with local gravitational dynamics strongly constrains
the values of $\eta$ and $k_c$ as well as 
the form of F near the transition acceleration (eqs.\ 48).  In the high
acceleration limit and {\it in the preferred frame}, 
the Lagrangian becomes that of 
a weakly coupled scalar field, as in Brans-Dicke theory with a 
large value of the Brans-Dicke parameter $\omega$.  However, the theory 
differs from the standard scalar-tensor theories 
in that the relation between the Einstein and physical metrics is
non-conformal.
Moreover, as is well-known from the measurement of
the CMB dipole anisotropy, we are not in the preferred frame:
the solar system is moving with a velocity of
370 km/s with respect to the cosmological frame; therefore, 
in addition to those relativistic effects associated with a weakly
coupled scalar field,
geophysical and orbital preferred frame effects
(Nordtvedt and Will 1972, Ni 1972) must also be present at some level.
These include diurnal solid earth tides, an annual variation of the
earth's rotation frequency and additional contributions to the anomalous
precession of planetary orbits.

For comparison with GR the magnitude of relativistic and 
preferred frame effects peculiar to any alternative theory  
are conveniently expressed in terms of the parameterized post-Newtonian (PPN)
formalism.  For the candidate theory the standard parameters   
can be evaluated following the procedure outlined
by Will (1993), and it is found that
$$\gamma = 1 \eqno(50a)$$
$$\beta = \Bigl({{1-\eta/2}\over {1+\eta/2}}\Bigr)^2 \eqno(50b)$$ 
$$\alpha_1 = {{-4\eta}\over{1+\eta/2}} \eqno(50c)$$
$$\alpha_2 = \alpha_3 = 0 \eqno(50d)$$
Those PPN parameters associated with the violation of energy-momentum
conservation are zero; i.e., the theory is ``semi-conservative'' 
in the terminology of Will (1993).

The value of $\gamma$ is the same as in GR implying an identical 
predicted deflection of light about the sun (not surprising since the 
theory was designed with this in mind) as well as identical predictions
for radar echo delay.  The parameter $\beta$ (= 1 in GR) 
enters into the expression
for anomalous relativistic precession of planetary orbits,
but the strongest expermental limit is provided by the lunar laser
ranging test of the equivalence principle (Dickey et al. 1994).  This
constrains $\beta < 10^{-4}$; therefore, from eq.\ 50b it must be the case
that $\eta<10^{-4}$.
 
The various preferred frame effects are expressed
in terms of the velocity of the solar system (or earth) 
with respect to the cosmological frame, {\bf w},
to second order in $w/c$, times various combinations of 
the three post-Newtonian parameters $\alpha_1$, $\alpha_2$,
and $\alpha_3$ 
(in GR all are zero).  Will and Nordtvedt (1973) demonstrated that for
all standard Lagrangian-based stratified theories 
(conformally flat on space-like
strata in the preferred Universal rest frame) it is the case that
$\alpha_2 = \alpha_3 = 0$, and if the light bending
is equivalent to that predicted by GR ($k = 2$ in
eq.\ 20), $\alpha_1 = -8$.  However, in the present case, 
where the physical metric is constructed from the Einstein metric
and not from the Minkowski
metric (eq.\ 7), the preferred frame effects are 
suppressed by roughly a factor of $4\eta$ (in the limit where $\eta$ becomes
very large, $\alpha_1 \rightarrow -8$ as in the standard stratified theories).
Combined solar system
data constrain $|\alpha_1|<4 \times 10^{-4}$ (Will 1993). 
A constraint on the strong-field equivalent of this parameter,
$\hat\alpha_1$,
implied by binary pulsar data, is $|\hat\alpha_1|<1.7\times 10^{-4}$ 
(Bell et al. 1996).  
But very recently it has been pointed out that current lunar ranging 
already constrains $\alpha_1$
at a level below $10^{-4}$ (M\"uller et al. 1996));  
therefore, if we take 
an experimental upper limit of $|\alpha_1| < 5\times 10^{-5}$ it must be
the case that $\eta < 1.25\times 10^{-5}$ if the proposed theory is
to be viable (i.e., in the limiting case $\alpha_1 = -5\times 10^{-5}$
which is consistent with the present limit of $-8 \pm 9 \times 10^{-5}$
determined by M\"uller et al. 1996).

While the non-detection of preferred frame effects at this level of
precision provide an upper limit on $\eta$,
the two requirements of inverse square attraction in the solar system
and of asymptotically flat, featureless galaxy rotation curves 
provide, in effect, a lower limit, as well as determining
the value of $k_s$.  The total attraction in the high acceleration limit is,
$$f_\odot = (1+\eta/2){{GM_\odot}\over {r^2}}, \eqno(51)$$ and this 
should extend at least to the orbit of Neptune to assure consistency
with the experimental result of Anderson et al. (1995).
Taking $a_o = 1.2\times 10^{-8}\, {\rm cm/s^2}$
as above,  this means that inverse square attraction should extend to
$(f_\odot/a_o)_{Nep} = 5.5\times 10^{4}$.  In other words, 
the high acceleration limit of the theory (eq.\ 48a) should apply down 
to a transition scalar acceleration of
$$k_c = f_s/a_o \leq (\eta/2)(f_\odot/a_o)_{Nep} \eqno(52)$$
But, as noted above, the prediction of flat featureless 
rotation curves as in Fig.\ 2
requires that $k_c$ be greater than 0.3.  This, combined with eq.\ 52,
and the upper limit on $\eta$ set by the experimental limit on 
$\alpha_1$, requires that $$0.30 < k_c < 0.35$$ and $$1.0\times 10^{-5}<
\eta <1.3\times 10^{-5}.$$  Thus the window of viability for this theory
is very small indeed.  In particular, the lower limit on $\eta$ implies that  
local preferred frame effects should soon be 
detectable at the level of $\alpha_1 \ge 4\times 10^{-5}$
if this theory is correct.

In a general sense, a stratified 
theory constructed from the Einstein metric, as opposed to the Minkowski
metric, can predict very weak local preferred frame effects because 
it is a two-field theory with a non-standard scalar field action:  
in addition to the scalar force there
is the usual Einstein-Newton force.  It is the scalar force which ties
the solar system to the cosmological frame;  the local tensor field is
not influenced by motion with respect to this frame. 
Because of the peculiar aquadratic scalar-field Lagrangian
it is this usual Einstein-Newton force (the first term
in eq.\ 13) which becomes dominant in the limit of large accelerations
(in the solar system or on the surface of the earth).
In effect, the preferred frame effects are suppressed by the factor
$f_s/f_n$, the ratio of the scalar force to the Newtonian force.
Thus the very same scalar-field Lagrangian which yields MOND phenomenology
on the scale of galaxies suppresses local preferred frame effects.

On the scale of galaxies, the theory is not
Newtonian, so it is inappropriate to speak of post-Newtonian parameters.
But, because the scalar force dominates the Einstein-Newton force on
this scale, the preferred frame effects should be present with their
full magnitude.  It is not clear that this could influence the structure
of galaxies.

With respect to the original binary pulsar, the
candidate theory, as a scalar-tensor theory, would lead to
the emission of dipole radiation in addition to the usual
quadrapole gravitational radiation.  However, because the scalar
field is so weakly coupled in the high-acceleration limit, it is
expected that the dipole radiation would also be suppressed
by a factor of $\eta$. 
Thus there is not likely to be a predicted
contradiction with the observed rate of orbital decay in the
binary pulsar; although this has not yet been worked out in detail.

There is one additional local scalar-tensor effect which cannot be
suppressed.  As in
any scalar-tensor theory there is a cosmic variation of the constant
of gravity.  In this case the magnitude of this effect is
$${\dot G\over G} = -{\dot\phi_o} \eqno(53) $$
For the toy theory considered here, with the use of eq.\ 43,
this becomes
$${\dot G\over G} = {2k_c} a_o/c \approx {{7.0\times 10^{-12}}} (a_o/
10^{-8} cm\,s^{-2})\,\,{\rm year^{-1}}. \eqno(54) $$
There is an similar expression in the context of PCG
(Sanders 1989) and this suggests that ${\rm {\dot G/G} \approx a_o/c}$ applies
to any cosmological effective theory for MOND based upon scalar-tensor
theory.  Determination of ${\rm \dot G/G}$ by ranging measurements 
are already at levels of precision
below $10^{-11}\,\, {\rm {year}^{-1}}$ (Will 1993); thus time
variation of G should also soon be detected if MOND is correct and
scalar-tensor theory is its basis.

In summary, aquadratic stratified theories of modified dynamics
are very strongly constrained by three
observational requirements:  the necessity of producing almost
perfect inverse square attraction in the solar system out to Neptune, 
the avoidance of detectable preferred frame effects at the location of 
the earth (at least at the present levels of experimental precision);
and the necessity of predicting MOND phenomenology on the scale of
galaxies and clusters of galaxies.
The toy theory described by eqs.\ 48 barely satisfies these 
requirements.  Values of $\eta$ much larger than $10^{-5}$ are ruled 
out by the present constraints on local preferred frame effects.
Smaller values of $\eta$ suppress preferred frame effects but at the
expense either of inverse square attraction in the solar system,
or of asymptotically flat, structureless galaxy rotation 
curves.  Changing the theory such that the scalar force falls 
more rapidly with
radius in the transition region between inverse-square and MOND attraction
(a BM function intermediate between eqs.\ 45 and 48) makes matters worse; 
such theories are already ruled out because they violate the constraints
on either local inverse square attraction or preferred frame effects.  
A scalar force which which actually increases with radius could work but
such a theory is impossible in the context of one-scalar aquadratic theory.
So in that sense, eqs.\ 48 describe a limiting case
for aquadratic stratified theories in which the scalar force decreases 
monotonically with radius; if this theory is not viable, then all such theories
are not viable.  

\section{Conclusions}

As first emphasized by Milgrom, 
the near numerical coincidence of the MOND acceleration 
parameter $a_o$ with $cH_o$ provides a very important clue to the
theoretical basis of MOND:  the cosmological background
affects local dynamics in a way which is closely
approximated by the MOND prescription.  
Such considerations clearly rule out GR because this
theory predicts no such direct cosmological influence of this magnitude
on local dynamics.  AQUAL, an unconventional scalar-tensor theory
for MOND (BM, Sanders 1986), is in no sense such an effective theory 
because $a_o$ is 
explicitly written in by hand and the theory has no cosmological limit.
PCG (Bekenstein 1988), as one of a class of scalar-tensor 
theories characterized by two
scalar fields coupled in the kinetic term of one of them, is
such an effective theory with the cosmic time derivative of the
matter-coupling field playing the role of $a_o$ (Sanders 1989).
However, AQUAL, PCG and all scalar-tensor theories
in which the field couples to matter as a conformal factor multiplying 
the Einstein metric fail to reproduce the observed deflection of light
by clusters of galaxies (Bekenstein and Sanders 1994).  

This problem
can be solved by reintroducing and generalizing the concept of conformal
coupling only on space-like strata of a preferred universal frame.
The traditional means of singling out such a preferred frame is through 
the introduction of a non-dynamical universal vector field
with the only non-zero component being the time component in the 
preferred cosmic frame. 
The essential new ingredient presented in this paper is that the
introduction of such a cosmic vector field can, at the same time,
both solve the light-bending problem of scalar-tensor theories and
also permit an aquadratic scalar field Lagrangian to be written 
without the explicit introduction of a new dimensional parameter
$a_o$.  That is to say, with this single new element
one can write a cosmological effective scalar-tensor theory
of MOND which also predicts the degree of the gravitational
deflection of light actually observed in cosmic gravitational lenses.

But it should be emphasized that this is not a traditional stratified
theory in that the physical metric is constructed from the Einstein
metric and not the Minkowski metric (eq.\ 7). 
The only {\it a priori} element is the vector field.
Therefore, Einstein's field equations are retained (with additional
source terms), and, in the particular
candidate theory considered here,  the
traditional Einstein-Newton force becomes dominant in the high acceleration
regime-- such theories are indistinguishable from GR to high precision 
on the scale of
the inner solar system and binary pulsar.  For accelerations comparable to
those prevailing in the inner solar system, the toy theory considered here
reduces to a weakly-coupled scalar tensor theory which, because of the
non-conformal relation between the two metrics, differs from
Brans-Dicke theory in that the predicted light
deflection and radar echo delay are precisely the same as in GR, although
preferred frame effects are present at some level. 
However, the dominance
of the Einstein-Newton force on this scale also implies that the
inevitable preferred frame effects are suppressed by a factor of
$4\eta$, where $\eta$ is a parameter of the theory 
that can be small but not arbitrarily
small.  Moreover, theories of this form make rather
precise predictions on the cosmic variation of the constant of gravity 
($\approx a_o/c$)
Of course, one could reasonably expect that any relativistic 
generalization
of MOND would lead to local deviations from the predictions of GR at
some level.  From this point-of-view the continued design of local
gravity tests with higher precision is a valuable activity.

The structure of this class of theories (eqs.\ 32) appears, at least 
superficially, to be similar to the aquadratic theories
of BM and of Sanders (1986). These
earlier AQUAL theories have been considered unphysical 
because of the predicted superluminal propagation of scalar waves and
the implied violation of causality (Bekenstein 1988).  
The same objection does not necessarily apply
to theories of this general type because the replacement of
the MOND function $\mu$ by a tensor $P^{\alpha\beta}$ (eq.\ 32a)
does, in fact, give the theory a different structure.
In general, the properties of stability and causality depend upon
the precise form of $F$ and the value of $\kappa$ (eq.\ 33).
Whether or not causality and stability
can be reconciled with
the cosmological origin of $a_o$ is the subject of a later paper on
AQUAL and two-scalar preferred frame theories.

The introduction of an {\it a priori} field is an unattractive element
in any theory.  It might well be that in a consistent theory the 
vector field must be dynamical--
a universal vector field coupled to gravity as in the theory of 
Will and Nordtvedt (1972), or
perhaps the normalized scalar field gradient itself as in Bekenstein's
(1992) disformal transformation.  Dynamical or not, the
vector singles out a preferred frame;  in more fully dynamical theories
the preferred frame effects might be further suppressed by the appearance
of cosmological matching parameters as in the generalized stratified theory
of Lee et al. (1974).
But it should be noted that the vector
field as written here does more than select a preferred frame;  it
also breaks the time-reversal invariance of gravitational physics in a 
fundamental way.  It literally is the arrow of time written in by hand.

From an observational point-of-view there clearly is a universal preferred
frame-- that in which the CMB dipole vanishes.  There is also a universal  
cosmic time which appears to possess a sense of direction not present in
the space-like dimensions of this frame.  One might speculate that cosmology
is described by a preferred-frame, time-irreversible theory of gravity
-- a stratified theory with long range interaction primarily
mediated by a scalar field--
while local gravitational dynamics, i.e., at accelerations higher than the
natural cosmic value of $cH_o$, are described by GR.  A
consequence of such a supposition is the necessity of scalar field dynamics
similar to that of the toy theory described here-- i.e., aquadratic 
scalar-tensor theory or an equivalent two-scalar theory like PCG.
That is to say, the reconciliation of preferred frame
cosmology with general relativistic local dynamics (very weak local 
preferred frame effects) would {\it require} the Modified Dynamics at 
low acceleration.  But such speculation is only suggested by the 
observational appearance of a preferred cosmic frame and, at present, 
has no justification in deeper theory.

In summary, the result of Bekenstein and Sanders (1994) on gravitational
lensing in the context of scalar-tensor theories (i.e., that the lensing   
mass of a cluster should be substantially less than the virial mass)
dealt a serious blow to such theories as a foundation for MOND.
The essential result here is that scalar-tensor theories can be cured of
this ailment
apparently only at the expense of rewriting them as stratified theories,
but then, the Lorentz Invariance of gravitational dynamics
is broken and local preferred frame effects
are inevitable.  The non-standard scalar-field Lagrangian 
giving rise to MOND phenomenology allows preferred
frame effects to be suppressed, but not by an arbitrary factor due to 
necessity of producing asymptotically flat spiral galaxy rotation curves
while satisfying the strict experimental
limits on deviations from inverse-square attraction in the
outer solar system.  The non-detection of preferred frame effects at
a level of a factor of three below current limits would not necessarily
rule out
MOND, but it would inflict serious damage on stratified scalar-tensor   
theories of MOND.  It is clear that
if one adopts the point-of-view that the mass discrepancy in
large astronomical systems is due to an incomplete understanding of gravity
rather than to dark matter, then consistency with phenomena ranging from
gravitational lensing by clusters of galaxies to galaxy rotation curves   
to planetary motion already imposes stringent conditions on acceptable
field theories.

\acknowledgments

I am very grateful to J.D. Bekenstein and M. Milgrom. Although they do  
not necessarily endorse everything in this paper, many of the
ideas here have emerged as a result of our numerous discussions and their
deep insight.  I also thank the Center for Microphysics and Cosmology
at the Hebrew University for support during a visit to Jerusalem in 1994.

\clearpage

\figcaption[ ]  {A log-log plot of the Newtonian force $\rm f_n$ 
and the scalar force $\rm f_s$ as a function of distance $\rm r$ from a
point mass M.  The scalar force is plotted (solid curves) 
for the theory described 
by eqs.\ 48 where $\eta = 1.25\times 10^{-5}$ and 
$k_c = f_s({\rm Neptune})/a_o = 0.34$.  
The force is given in units of $\rm a_o$ and the radius
in terms of $r_o = \sqrt{GM/a_o}$. 
The asymptotic behavior of the scalar force is consistent with
perfect inverse square attraction in the solar system to the orbit
of Neptune, indicated by the vertical dashed line (where $a_o=1.2
\times 10^{-8}$ ${\rm cm/s^2}$), but approaches 
1/r at low accelerations (the MOND limit).
Smaller values of $\eta$ further surpress preferred frame effects
but at the expense of inverse square attraction within the orbit
of Neptune.  Smaller values of $k_c$ increase the extent of inverse
square attraction (for a given value of $\eta$) but at the expense
of asymptotically flat rotation curves.
Also shown by the dashed curve is the
circular velocity $V_c$  in units of the MOND asymptotic velocity
$(GMa_o)^{1/4}$.  This is Keplerian inside the transition radius
but approaches one asymptotically. \label{fig1}}  

\figcaption[ ] {The predicted rotation curves of spherical galaxies
having an exponential density distribution.  The solid curve shows the
total rotational velocity as a function of radius.  This includes,
in addition to the usual Newtonian force,  the
scalar force which is derived from the toy theory described by eqs.\ 48;
i.e., the theory is identical to that giving the force about a 
point mass shown in Fig.\ 1.
The dashed curve is the rotation curve resulting from the Newtonian force
alone.  The masses of the galaxies (in solar units) and the exponential
scale lengths are indicated on the figure.  The rotation curves are seen
to be asymptotically featureless and flat;  the asymptotic velocity scales
as the one-fourth power of the mass as in MOND. \label{fig2}}

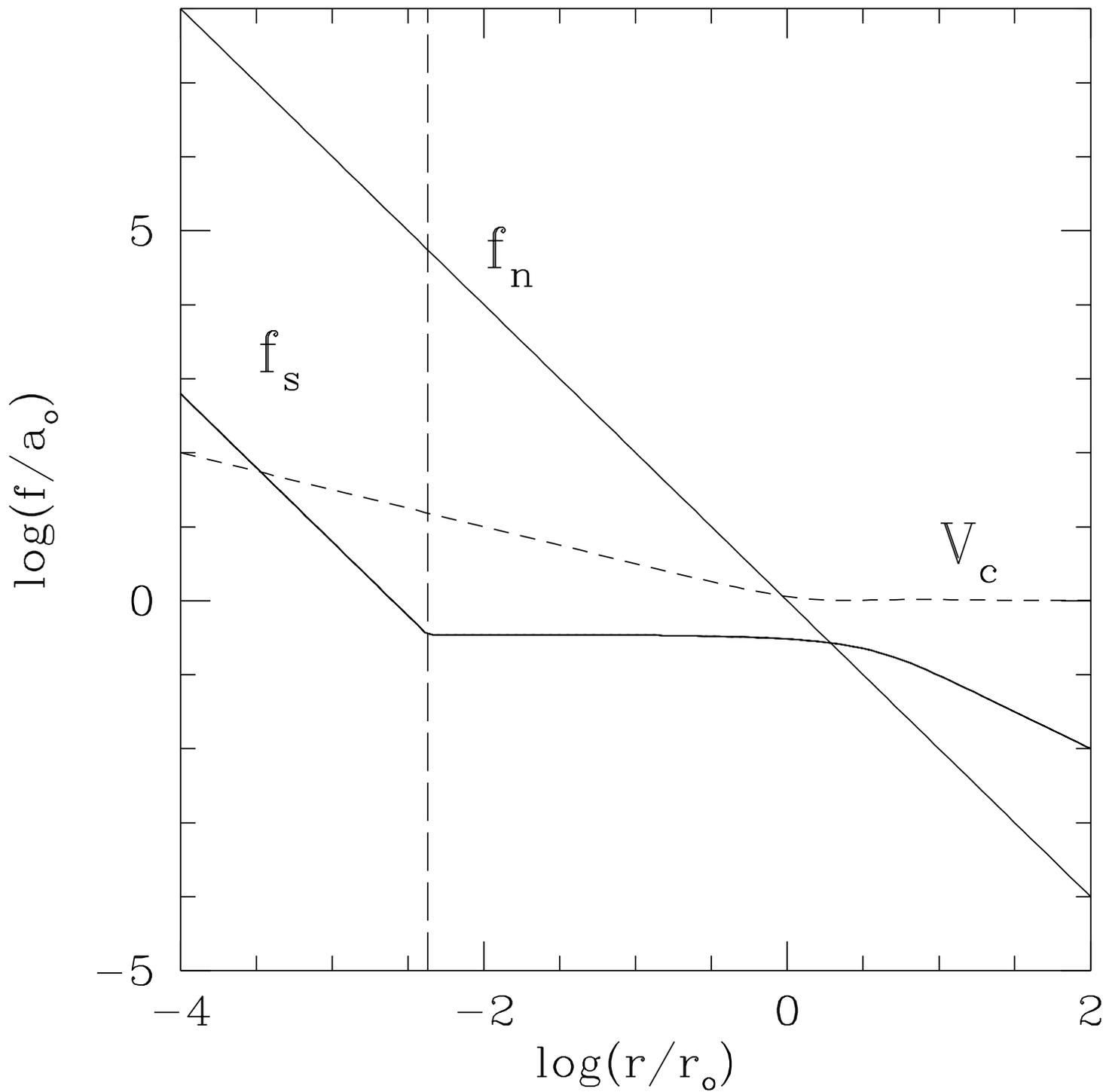
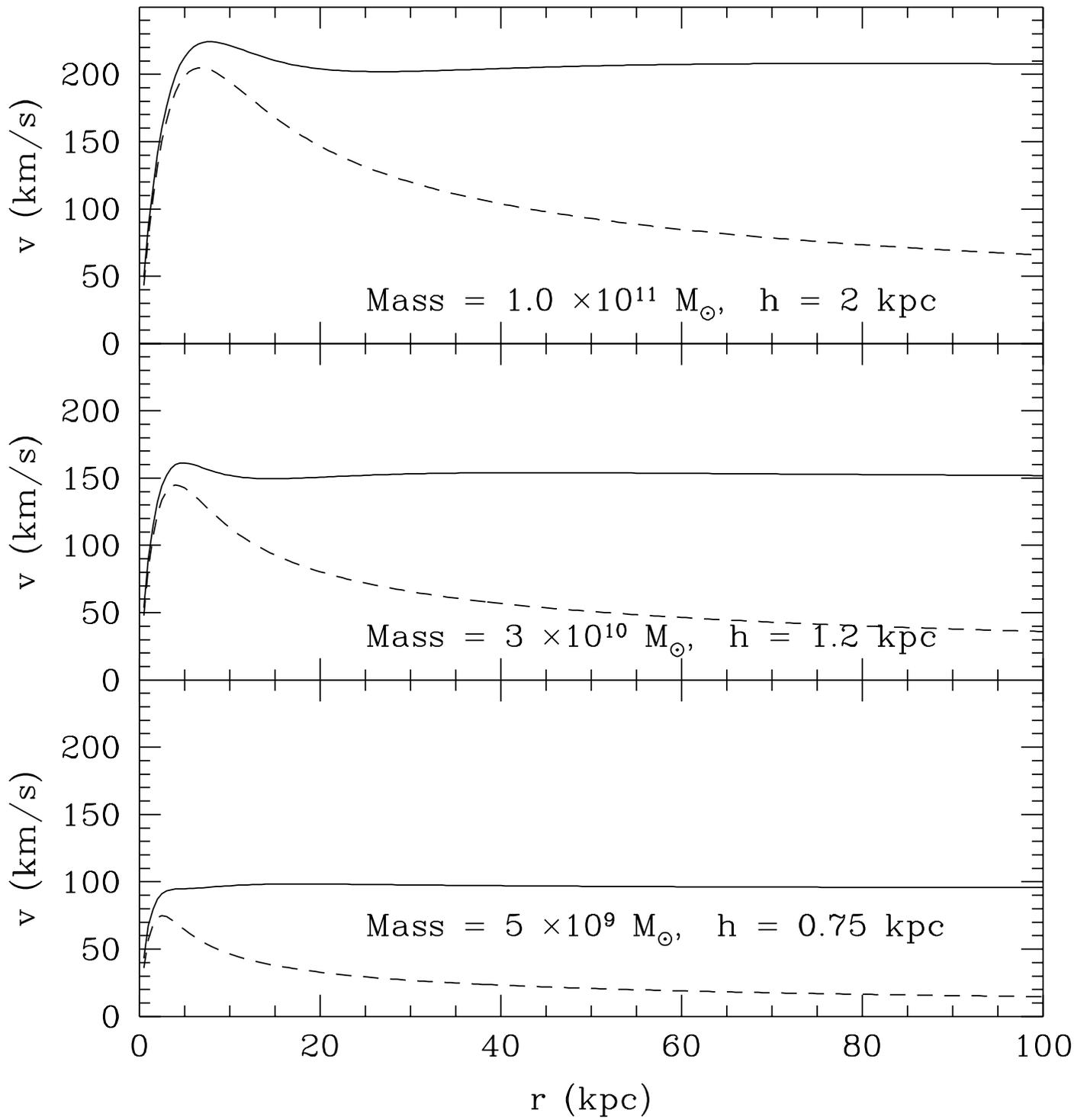
\end{document}